\documentclass[11pt]{article}
\usepackage{mathrsfs}
\usepackage{amsmath}
\usepackage{amsfonts}
\usepackage{amssymb, amsmath, cite}
\usepackage{color}
\usepackage{graphicx}

\newcommand\be{\begin{equation}}
\newcommand\ee{\end{equation}}
\newcommand\ber{\begin{eqnarray}}
\newcommand\eer{\end{eqnarray}}
\newcommand\berr{\begin{eqnarray*}}
\newcommand\eerr{\end{eqnarray*}}
\newcommand\bea{\begin{eqnarray}}
\newcommand\eea{\end{eqnarray}}

\newcommand{\bfR}{{\Bbb R}}

\newcommand{\dd}{\mbox{d}}\newcommand{\D}{\cal O}
\newcommand{\e}{\mbox{e}}
\newcommand{\pa}{\partial}\newcommand{\om}{\omega}
\newcommand{\vep}{\varepsilon}

\newcommand{\nn}{\nonumber}

\newcommand\lb{\label}
\newcommand\eq{\eqref}

\title{Exact Solutions to the Nonlinear\\ Governing Equations of the Born--Infeld Theory}

\author{
Yisong Yang\footnote{Email address: yisongyang@nyu.edu}\\Courant Institute of Mathematical Sciences\\ New York University\\New York, New York 10012, USA
}

\date{}

\begin{document}

\maketitle

\begin{abstract}
Exact  finite-energy solutions to the nonlinear governing equations of the Born--Infeld theory of electrodynamics,  describing continuous distributions of electric, magnetic, and
dyonic charge sources, in both classical and generalized settings, are constructed explicitly. In particular, it is shown that, the finiteness of the total prescribed charges leads to the finiteness of
the total energy, of the electromagnetic field system. As a by-product, this result resolves a puzzle arising in the dyonic point charge distribution situation where energy divergence inevitably occurs.
\medskip

{\flushleft {Keywords:} Born--Infeld theory,  nonlinear partial differential equations, exact  solutions, free electric and magnetic charges, finite charge, finite-energy, dyonic solutions.
\medskip

\medskip

}
\end{abstract}

\section{Introduction}

In this work, we explicitly construct, for the first time, exact solutions to the nonlinear governing equations in the classical electrodynamics theory of  Born and Infeld and its generalizations
subject to the presence of {\em arbitrarily distributed continuous static charges}, which can be purely electric, magnetic, or both. That is,  the charge distribution in our construction may also be allowed to be dyonic or dually electric and
magnetic.

It is well known that, initiated in 1933,  Born and Infeld developed a geometrized theory of electromagnetism to resolve the energy divergence problem associated with an electric point charge \cite{B1,B2,B3,B4}, and that,
more recently, this theory and its various generalized forms have found applications  in many areas of modern theoretical physics including superstrings \cite{FT,Ts2,Tsey}  and branes \cite{CM,Gibbons,Ts1}, charged black holes \cite{AG1,AG2,K1,K2,Yang1,Yang2,Yang3},
and cosmology \cite{Jana,Kam,Nov,Yang1,Yang3}. In particular, the article \cite{JHOR} offers a survey on a very rich range of modified gravity theories inspired by \cite{B1,B2,B3,B4}
aimed at tackling some theoretical issues unsettled in the framework of classical theory.

The governing equations arising in the Born--Infeld type theories are notoriously difficult due to their highly challenging nonlinear structures.
In fact, in the original work of Born and Infeld, only a radially symmetric solution describing a singly-centered electric charge is constructed which is used to show how to
resolve the energy divergence problem of a point charge. In fact, all subsequent studies based on the governing equations associated with the theories developed have been
made  using radially symmetric solutions productively.

In order to understand how charged particles interact with each other
electromagnetically in the Born--Infeld type theories, it will be important to construct exact solutions in these settings realizing  multiply-centered 
 source distributions. Such a task is accomplished in the recent work \cite{Yang-AOP} in which exact solutions of the Born--Infeld equations, and their extensions,
realizing {\em multicentered static point charge distributions} are constructed explicitly for the first time. In the present work, we extend the method in \cite{Yang-AOP} to
obtain exact solutions realizing arbitrarily distributed continuous charged sources explicitly and systematically.

Our method is based on recognizing that the Born--Infeld theory type electrodynamics equations in their static situation are of the form of 
 the generalized covariant Maxwell equations of Schwinger \cite{Sch1,Sch2,Sch3}:
\be\lb{1}
\nabla\times{\bf E}=-{\bf j}_m,\quad \nabla\times{\bf H}={\bf j}_e,\quad \nabla\cdot{\bf D}=\rho_e,\quad \nabla\cdot{\bf B}=\rho_m,
\ee
describing the interaction of electric and magnetic fields $\bf E$ and $\bf B$, and electric displacement field $\bf D$ and magnetic intensity field $\bf H$, which in the simplest Born--Infeld setting are 
formally related through a pair of nonlinear constitutive equations,
\be\lb{2}
{\bf D}=\vep({\bf E},{\bf B}){\bf E},\quad {\bf B}=\mu({\bf E},{\bf B}){\bf H},
\ee
in the presence of electric and magnetic charge densities $\rho_e$ and $\rho_m$, and current densities ${\bf j}_e$ and ${\bf j}_m$. For a statically charged system, the charge densities
$\rho_e$ and $\rho_m$ may be prescribed which determine the fields $\bf D$ and $\bf B$ by the last two equations in \eq{1}. With so determined $\bf D$ and $\bf B$, we solve for
$\bf E$ from the first equation in \eq{2} which is an implicit equation. Inserting $\bf E$ into the second equation, we find $\bf H$. In this approach, the pairs of  the fields $\bf D$, $\bf B$
and $\bf E$, $\bf H$ play rather different roles. The former are directly related to the prescribed charge densities, $\rho_e$ and $\rho_m$, and the latter are induced through \eq{2}, which
then generate the magnetic and electric current densities ${\bf j}_m$ and ${\bf j}_e$ through the first two equations of the system \eq{1} which are present to balance the multiply-centered
charges as in the point charge situation \cite{Yang-AOP}. In this spirit, $\bf E$ and $\bf H$ also generate the free electric and magnetic charge densities
$\rho_{f,e}$ and $\rho_{f,m}$  through the expressions
\be\lb{3}
\nabla\cdot{\bf E}=\rho_{f,e},\quad \nabla\cdot{\bf H}=\rho_{f,m},
\ee
following \cite{B4}. Physically, $\rho_e$ and $\rho_m$ are theoretically prescribed and $\rho_{f,e}$ and $\rho_{f,m}$  ``laboratorily measured" or what available for measurement. It is
shown by Born and Infeld in \cite{B4} in the point-charge situation that the prescribed charge and free charge agree. A by-product of our study here is that this agreement is true in general.
That is, we establish the identities
\be\lb{4}
Q_e\equiv \int_{\bfR^3} \rho_e\,\dd x=\int_{\bfR^3} \rho_{f,e}\,\dd x\equiv Q_{f,e},\quad Q_m\equiv\int_{\bfR^3} \rho_m\,\dd x=\int_{\bfR^3} \rho_{f,m}\,\dd x\equiv Q_{f,m},
\ee
in {\em all} general situations. It should be noted that such a result is not generally valid in the point charge situation \cite{Yang-AOP} involving dual charges. That is, when both electric and
magnetic charges are present, it may occur that \eq{4} is violated. More precisely, in such a dyonic situation, the free charges may be given in mixed forms of the types:
\be\lb{5}
Q_{f,e}=Q_e-Q_m,\quad Q_{f,m}=Q_m-Q_e,
\ee
which appear as an indicator of local energy divergence associated with the point charges.  In the continuous situation, there is no more local energy divergence so that \eq{5} will never
occur.  This result resolves the puzzle associated with the concurrence of the charge mixing laws \eq{5} and energy divergence in the point charge situation \cite{Yang-AOP}.

Here is an outline of the presentation of this work. In Section \ref{sec2}, we review the Born--Infeld theory and reformulate it in view of Schwinger's covariant Maxwell equations. In Section \ref{sec3}, we
treat the simplest situation which lays the foundation of the rest of the work.
 In Sections \ref{sec4} and \ref{sec5}, we deal with the general situation of the classical Born--Infeld theory. In Section \ref{sec6}, we
extend our study to the generalized Born--Infeld theories. In Section \ref{sec7}, we present a few concrete model examples as further illustrations of our method.
 In Section \ref{sec8}, we summarize our results.

\section{Nonlinear electrodynamics and governing equations with sources}\lb{sec2}
\setcounter{equation}{0}

Consider the Minkowski spacetime $\bfR^{3,1}$ with the metric tensor $\eta_{\mu\nu}=\mbox{diag}\{1,-1,-1,-1\}$. Let $A_\mu$ be a real-valued vector gauge field and  $F_{\mu\nu}$ the induced electromagnetic tensor field given by 
\be\lb{F}
F_{\mu\nu}=\pa_\mu A_\nu-\pa_\nu A_\mu, \quad \mu,\nu=0,1,2,3.
\ee
Use $\epsilon^{\mu\nu\alpha\beta}$ to denote the usual Kronecker skewsymmetric symbol such that the Hodge dual of $F_{\mu\nu}$ is defined by 
$\tilde{F}^{\mu\nu}=\frac12\epsilon^{\mu\nu\alpha\beta}F_{\alpha\beta}$ which gives rise to the corresponding electromagnetic duality:
\bea
(F^{\mu\nu})&=&\left(\begin{array}{cccc}0&-E^1&-E^2&-E^3\\E^1&0&-B^3&B^2\\E^2&B^3&0&-B^1\\E^3&-B^2&B^1&0\end{array}\right),\lb{x2.3a}\\
(\tilde{F}^{\mu\nu})&=&\left(\begin{array}{cccc}0&-B^1&-B^2&-B^3\\B^1&0&E^3&-E^2\\B^2&-E^3&0&E^1\\B^3&E^2&-E^1&0\end{array}\right),\lb{x2.3b}
\eea
where  ${\bf E}=(E^i)$ and  ${\bf B}=(B^i)$ are the associated, underlying, electric and magnetic fields, respectively.
With this preparation of notation,  the free Lagrangian action density of the classical Born--Infeld theory of nonlinear electrodynamics reads \cite{B4,Yang2,Yang3,Kr1}
\bea
&&{\cal L}=\frac1{\beta}\left(1-\sqrt{1-2\beta s}\right),\lb{1.1}\\
&&s=-\frac14F_{\mu\nu}F^{\mu\nu}+\frac{\kappa^2}{32}\left(F_{\mu\nu}\tilde{F}^{\mu\nu}\right)^2,\lb{x2.2}
\eea
where 
the parameter $\beta>0$ relates to the usual Born parameter, and $\kappa>0$ is an electromagnetic 
coupling parameter that directly mixes the interaction of the fields ${\bf E}$ and ${\bf B}$. Later we shall see that the quantity $s$ given in \eq{x2.2} plays a crucial role in a systematic construction 
of exact solutions of the equations no matter how complicated the equations become.

When a source 4-current $j^\mu=(\rho_e,{\bf j}_e)$, with $\rho_e$ and ${\bf j}_e$ the electric charge and current densities, respectively, is present, we need to couple $A_\mu$ and
$j_\mu$ so that the action density \eq{1.1} assumes the modified form
\be\lb{x2.4}
{\cal L}=\frac1{\beta}\left(1-\sqrt{1-2\beta s}\right)-A_\mu j^\mu.
\ee
Varying the gauge field $A_\mu$  in \eq{x2.4} then leads to its Euler--Lagrange equations
\bea
\pa_\mu P^{\mu\nu}&=&j^\nu,\lb{x2.5}\\
P^{\mu\nu}&=&\frac1{\sqrt{1-2\beta s}}\left(F^{\mu\nu}-\frac{\kappa^2}4(F_{\alpha\beta}\tilde{F}^{\alpha\beta})\tilde{F}^{\mu\nu}\right),\lb{x2.6}
\eea
where the tensor field $P^{\mu\nu}$ is associated with the induced electric displacement field $\bf D$ and magnetic intensity field $\bf H$ through the matrix
\be\lb{x2.7}
(P^{\mu\nu})=\left(\begin{array}{cccc}0&-D^1&-D^2&-D^3\\D^1&0&-H^3&H^2\\D^2&H^3&0&-H^1\\D^3&-H^2&H^1&0\end{array}\right),
\ee
which is in a parallel spirit to that of the matrix \eq{x2.3a}.
With \eq{x2.3a} and \eq{x2.3b}, we see that the quantity $s$ given by \eq{x2.2} reads
\be
s=\frac12({\bf E}^2-{\bf B}^2)+\frac{\kappa^2}2({\bf E}\cdot{\bf B})^2,\lb{1.2}
\ee
and that \eq{x2.6} relates the fields ${\bf D}$, $\bf H$  to $\bf E$, $\bf B$ through the equations 
\bea
&&{\bf D}=\frac1{\sqrt{1-2\beta s}}({\bf E}+\kappa^2 [{\bf E}\cdot{\bf B}]{\bf B}),\lb{1.3}\\
&&{\bf H}=\frac1{\sqrt{1-2\beta s}}({\bf B}-\kappa^2 [{\bf E}\cdot{\bf B}]{\bf E}),\lb{1.4}
\eea
which may well be considered as a set of constitutive equations of the theory with field-dependent electric permittivity and magnetic permeability coefficients.
It should be emphasized that, since these equations are simply a realization of \eq{x2.6}, they are source independent.

First, in view of \eq{x2.5} and the matrix \eq{x2.7}, we arrive at 
\be\lb{x2.11}
-\frac{\pa{\bf D}}{\pa t}+\nabla\times{\bf H}={\bf j}_e,\quad \nabla\cdot{\bf D}=\rho_e,
\ee
which are the Amper\'{e} law and Coulomb law, respectively, in the Born--Infeld setting. 

Next, from the definition of $F_{\mu\nu}$ given in \eq{F}, we have the Bianchi identity
\be\lb{x2.12}
\pa^\gamma F^{\mu\nu}+\pa^\mu F^{\nu\gamma}+\pa^\nu F^{\gamma\mu}=0, \quad\mbox{or}\quad \pa_\mu \tilde{F}^{\mu\nu}=0,
\ee
which in view of \eq{x2.3a} and \eq{x2.3b}, splits into two equations as before:
\be\lb{x2.13}
\frac{\pa{\bf B}}{\pa t}+\nabla\times{\bf E}={\bf 0},\quad\nabla\cdot{\bf B}=0.
\ee
These equations are the Faraday law and Gauss law, respectively, in the Born--Infeld theory formalism \cite{B1,B2,B3,B4}. 

Coupled through the nonlinear constitutive relations  \eq{1.3} and \eq{1.4},
the equations \eq{x2.11} and \eq{x2.13} assume the form
of the classical Maxwell equations apparently subject to the same electromagnetic asymmetry, meaning that electricity and magnetism appear on different footings so that electric charge and
current densities are present but magnetic ones are absent. 

On the other hand, in the vacuum situation free of external sources, there holds the electromagnetic symmetry due to the duality structure realized by \eq{x2.3a} and \eq{x2.3b}.
It was
Dirac \cite{Dirac} and Schwinger \cite{Sch1,Sch2,Sch3} who showed that the broken electromagnetic symmetry or duality in presence of external sources can be restored
if we add magnetic charge and current densities, say $\rho_m$ and ${\bf j}_m$, and subject them to transform  following the corresponding duality relation:
\be
\rho_e\mapsto\rho_m,\quad {\bf j}_e\mapsto {\bf j}_m,\quad \rho_m\mapsto-\rho_e,\quad {\bf j}_m\mapsto -{\bf j}_e,
\ee
such that \eq{x2.13} becomes
\be\lb{xx2.17}
\frac{\pa{\bf B}}{\pa t}+\nabla\times{\bf E}=-{\bf j}_m,\quad\nabla\cdot{\bf B}=\rho_m.
\ee

In this context,
the equations \eq{x2.11} and \eq{xx2.17}  are formally the generalized Maxwell equations \cite{Bla,Sch1,Jackson,Sin,Singleton,Mig,Milton} for which the fields ${\bf E},{\bf B}$ and ${\bf D},{\bf H}$ obey
the usual linear constitutive equations
\be\lb{xx2.18}
{\bf D}=\vep{\bf E},\quad {\bf B}=\mu{\bf H},
\ee
with $\vep$ and $\mu$ the dielectrics and permeability coefficients, which are constants in free space such that $\frac1{\sqrt{\vep\mu}}=c=1$ gives us the normalized speed of light.

Although this formalism of the Maxwell equations breaks the Bianchi identity resulting in the invalidity of the variational principle, the elegant electromagnetic duality structure led Dirac \cite{Dirac}
and Schwinger \cite{Sch1,Sch2,Sch3}  (see also
 Zwanziger \cite{Z}) to conceptualize the notions of monopoles and dyons, which have become fundamental constructs in theoretical physics.

In the context of the Born--Infeld equations, the constitutive equations \eq{xx2.18} are replaced by \eq{1.3} and \eq{1.4} or
more explicitly by the vector-matrix equation
\be\lb{1.9}
\left(\begin{array}{c}{\bf D}\\{\bf B}\end{array}\right)=\Sigma({\bf E},{\bf B})\left(\begin{array}{c}{\bf E}\\{\bf H}\end{array}\right),\quad
 \Sigma({\bf E},{\bf B})\equiv\left(\begin{array}{cc}\frac{1+\kappa^4({\bf E}\cdot{\bf B})^2}{\sqrt{1-2\beta s}}& \kappa^2({\bf E}\cdot{\bf B})\\ \kappa^2({\bf E}\cdot{\bf B})&\sqrt{1-2\beta s}\end{array}\right),
\ee
which becomes diagonal of the form \eq{2} in the limiting situation $\kappa=0$ \cite{B1,B2,B3,B4}, so that the property $\det(\Sigma({\bf E},{\bf B}))=1$ realizes the normalized speed of light condition. Therefore, following the idea of Dirac \cite{Dirac} and Schwinger \cite{Sch1,Sch2,Sch3},
the Born--Infeld equations that enjoy the full electromagnetic duality or symmetry
are of the form
\bea
\nabla\cdot{\bf D}&=&\rho_e,\lb{1.7}\\
\nabla\cdot{\bf B}&=&\rho_m,\lb{1.8}\\
\frac{\pa{\bf B}}{\pa t}+\nabla\times{\bf E}&=&-{\bf j}_m,\lb{1.5}\\
-\frac{\pa{\bf D}}{\pa t}+\nabla\times{\bf H}&=&{\bf j}_e,\lb{1.6}
\eea
subject to applied electric and magnetic charge densities $\rho_e$ and $\rho_m$ and current
densities ${\bf j}_e$ and ${\bf j}_m$, respectively.
As remarked earlier,  a nonvanishing 
$\rho_m$ or ${\bf j}_m$ breaks the Bianchi identity, in view of \eq{x2.12} or \eq{x2.13}, thereby a gauge-field formalism,
as in the studies of Dirac \cite{Dirac} and Schwinger \cite{Sch1,Sch2,Sch3}.

We are interested in static solutions to \eq{1.7}--\eq{1.6} and we have shown in \cite{Yang-AOP} that the presence of multicentered point charges leads to the presence of currents
in order to achieve a static balance of the charges. Specifically, we have the following scenarios.

\begin{enumerate}

\item[(i)] In the multicentered point-charge situation realizing prescribed electric and magnetic charge distributions, the associated exact solutions necessarily give rise to
non-vanishing magnetic and electric currents, respectively. These currents disappear only in  singly-centered situations where radial symmetry prevails, but are present in
multicentered situations where radial symmetry is broken, in order to maintain a balanced equilibrium state of the point charge system.

\item[(ii)] When $\kappa=0$, the charge mixing laws expressed by \eq{5} hold and energy divergence takes place for dyonically charged solutions as a result of local
energy divergence of point charges.

\item[(iii)] When $\kappa>0$, the charge mixing laws expressed by \eq{5} give way to the charge laws stated in \eq{4} and total energies are all finite for all dyonically charged solutions as a result 
that the parameter $\kappa$ plays the role that switches off local
energy divergence associated with point dyonic charges.

\end{enumerate}

The main contribution of the present work is an explicit construct of exact solutions of these equations where $\rho_e$ and $\rho_m$ are continuous functions
realizing arbitrarily distributed electric, magnetic, and dyonic charge distributions. Since continuity of the charge distribution removes local energy divergence associated with
concentrated point charges, all solutions will be shown to carry finite energies and enjoy the exact charge laws expressed in \eq{4}, which resolves the puzzle \eq{5} concurring 
with energy divergence established in \cite{Yang-AOP} for dyonic cases.

\section{Explicit solutions and free charges when $\kappa=0$ in classical theory}\lb{sec3}
\setcounter{equation}{0}

Out of simplicity and independent interest, we first consider the situation where $\kappa=0$ of the classical theory \eq{1.1}--\eq{x2.2}. It should be noted that the Lagrangian in this limiting situation was originally
proposed by Born himself in \cite{B1,B2} without exploring its special relativity origin,
which was subsequently carried out in \cite{B3,B4} in Born's joint studies with Infeld. 
In this situation, \eq{1.3} and \eq{1.4} become
\bea
&&{\bf D}=\frac{\bf E}{\sqrt{1-\beta ({\bf E}^2-{\bf B}^2)}},\lb{b2.1}\\
&&{\bf H}=\frac{\bf B}{\sqrt{1-\beta ({\bf E}^2-{\bf B}^2)}}.\lb{b2.2}
\eea
From \eq{b2.1} alone, we can derive the identity
\be\lb{b2.3}
1-\beta({\bf E}^2-{\bf B}^2)=\frac{1+\beta{\bf B}^2}{1+\beta{\bf D}^2}.
\ee
Using \eq{b2.3} in \eq{b2.1}--\eq{b2.2}, we obtain the relations
\bea
&&{\bf E}={\bf D}\sqrt{\frac{1+\beta{\bf B}^2}{1+\beta{\bf D}^2}},\lb{b2.4}\\
&&{\bf H}={\bf B}\sqrt{\frac{1+\beta{\bf D}^2}{1+\beta{\bf B}^2}},\lb{b2.5}
\eea
which express $\bf E$ and $\bf H$ in terms of $\bf D$ and $\bf B$, which in turn are determined by \eq{1.7} and \eq{1.8}, respectively. In other words, we have seen that the system of
the equations \eq{1.7}--\eq{1.6}, in static limit, of course, is explicitly solved in terms of the prescribed electric and magnetic charge densities $\rho_e$ and $\rho_m$.

We now proceed to compute the free charges. The prescribed electric and magnetic charges defined by \eq{1.7} and \eq{1.8}, are given by the total integrals of $\rho_e$ and $\rho_m$,
 respectively, which may also be represented by the following surface integrals at the spatial infinity,
\bea
Q_e&=&\lim_{R\to\infty}\int_{|x|=R}{\bf D}\cdot\dd {\bf S},\lb{b2.8}\\
Q_m&=&\lim_{R\to\infty}\int_{|x|=R}{\bf B}\cdot\dd {\bf S},\lb{b2.9}
\eea
in view of the divergence theorem. The convergence of \eq{b2.8} and \eq{b2.9} leads to the assumption
\be
{\bf D}=\mbox{O}(r^{-2}),\quad {\bf B}=\mbox{O}(r^{-2}),\quad r=|x|\gg1,\lb{b2.10}
\ee
which is consistent with Coulomb's law as well to ensure the finiteness of the charges. This assumption will be observed in our subsequent development of the subject. Note that the equations
of the type \eq{1.7} or \eq{1.8} may be solved using standard potential theory \cite{Evans,GT,SY}.
Following \cite{B4}, define the free electric charge density $\rho_{f,e}$ to be
\be\lb{b2.11}
\rho_{f,e}=\nabla\cdot {\bf E}.
\ee
As a consequence of \eq{b2.8}, \eq{b2.4}, and \eq{b2.10}, we can now calculate the total free electric charge from \eq{b2.11}:
\bea\lb{xx3.12}
Q_{f,e}&=& \lim_{R\to\infty}\int_{|x|=R}{\bf E}\cdot\dd {\bf S}\nn\\
&=&\lim_{R\to\infty}\int_{|x|=R}{\bf D}\cdot\dd {\bf S}+ \lim_{R\to\infty}\int_{|x|=R}({\bf E}-{\bf D})\cdot\dd {\bf S}\nn\\
&=&Q_e+\lim_{R\to\infty}\int_{|x|=R}\frac{\beta({\bf B}^2-{\bf D}^2){\bf D}}{1+\beta{\bf D}^2+\sqrt{(1+\beta{\bf D}^2)(1+\beta{\bf B}^2)}}\cdot\dd {\bf S}\nn\\
&=&Q_e.
\eea
Similarly, we obtain the total free magnetic charge generated from the free magnetic charge density similarly defined \cite{B4} by the equation 
\be
\rho_{f,m}=\nabla\cdot {\bf H},
\ee
given by
\bea\lb{xx3.14}
Q_{f,m}&=&\int_{\bfR^3}\rho_{f,m}\,\dd x=\lim_{R\to\infty}\int_{|x|=R}{\bf B}\cdot\dd {\bf S}+ \lim_{R\to\infty}\int_{|x|=R}({\bf H}-{\bf B})\cdot\dd {\bf S}\nn\\
&=&Q_m- \lim_{R\to\infty}\int_{|x|=R}\frac{\beta({\bf B}^2-{\bf D}^2){\bf B}}{1+\beta{\bf B}^2+\sqrt{(1+\beta{\bf D}^2)(1+\beta{\bf B}^2)}}\cdot\dd {\bf S}\nn\\
&=&Q_m,
\eea
in view of \eq{b2.5}, \eq{b2.9}, and \eq{b2.10}. The results \eq{xx3.12} and \eq{xx3.14} confirm \eq{4}.

The total energy of the dyonic system will be considered later jointly together with the general situation $\kappa>0$ in Section \ref{sec5}.

\section{Explicit solutions and free charges when $\kappa$ is arbitrary}\lb{sec4}
\setcounter{equation}{0}

Using the idea of Section \ref{sec3} in the simplest situation $\kappa=0$, we extend our work to the general situation of
an arbitrary $\kappa$ where the dyonic charge sources are given by a pair of continuous functions $\rho_e$ and $\rho_m$ as stated in \eq{1.7} and \eq{1.8}.
More precisely, we are to resolve $\bf E$ and $\bf H$ in terms of $\bf D$ and $\bf B$ which are determined by \eq{1.7} and \eq{1.8} and considered known.
For this purpose, first, using \eq{1.2}, multiplying both sides of \eq{1.3} by $\bf B$,  squaring it, and then squaring both sides of \eq{1.3}, we arrive at the relations
\bea
({\bf B}\cdot{\bf D})^2(1-\beta[{\bf E}^2-{\bf B}^2+\kappa^2({\bf E}\cdot{\bf B})^2])&=&({\bf E}\cdot{\bf B})^2(1+\kappa^2{\bf B}^2)^2,\lb{3.1}\\
{\bf D}^2(1-\beta[{\bf E}^2-{\bf B}^2+\kappa^2({\bf E}\cdot{\bf B})^2])&=&{\bf E}^2+\kappa^2 (2+\kappa^2{\bf B}^2)({\bf E}\cdot{\bf B})^2,\lb{3.2}
\eea
where ${\bf E}^2$ and $({\bf E}\cdot{\bf B})^2$ are regarded as unknowns.
Solving  this system, we obtain
\bea
{\bf E}^2&=&\frac{(1+\beta{\bf B}^2)({\bf D}^2 +\kappa^2[2+\kappa^2{\bf B}^2][{\bf B}\times{\bf D}]^2)}{(1+\kappa^2{\bf B}^2)(1+\beta{\bf D}^2+\kappa^2{\bf B}^2+
\beta \kappa^2[{\bf B}\times{\bf D}]^2)},\lb{3.3}\\
({\bf E}\cdot{\bf B})^2&=&\frac{({\bf B}\cdot{\bf D})^2(1+\beta{\bf B}^2)}{(1+\kappa^2{\bf B}^2)(1+\beta{\bf D}^2+\kappa^2{\bf B}^2+
\beta \kappa^2[{\bf B}\times{\bf D}]^2)},\quad\lb{3.4}
\eea
where we have used the Lagrange identity 
$
({\bf B}\times{\bf D})^2={\bf B}^2{\bf D}^2-({\bf B}\cdot{\bf D})^2
$
to suppress the expressions on the right-hand sides of \eq{3.3} and \eq{3.4}.
Inserting \eq{3.3} and \eq{3.4} into \eq{1.3}, we obtain the electric field $\bf E$ explicitly:
\bea
{\bf E}&=&\sqrt{1-2\beta s}\,{\bf D}-\kappa^2({\bf E}\cdot{\bf B}){\bf B}\nn\\
&=&\sqrt{1-2\beta s}\left({\bf D}-\frac{\kappa^2({\bf B}\cdot{\bf D})}{1+\kappa^2{\bf B}^2}\,{\bf B}\right)\nn\\
&=&\frac{\sqrt{1+\beta{\bf B}^2}\left([1+\kappa^2{\bf B}^2]{\bf D}-\kappa^2[{\bf B}\cdot{\bf D}]{\bf B}\right)}{\sqrt{1+\kappa^2{\bf B}^2}\sqrt{1+\beta{\bf D}^2+\kappa^2{\bf B}^2
+\beta\kappa^2 ({\bf B}\times {\bf D})^2}}.\lb{3.6}
\eea
Although this formula looks rather complicated, it gives the dependence of $\bf E$ on $\bf D$ and $\bf B$ explicitly and clearly. In particular, it allows us to read off the sharp result
\be\lb{b3.7}
{\bf E}-{\bf D}=\mbox{O}(r^{-6}),\quad r=|x|\gg1,
\ee
from the asymptotic property \eq{b2.10} stated for the fields $\bf D$ and $\bf B$. This property is crucial for us to show below that the free electric charge is the same as the prescribed one.

Moreover, inserting \eq{3.3}, \eq{3.4}, and \eq{3.6} into \eq{1.4}, and after a lengthy simplification, we see that the magnetic intensity field $\bf H$ can be expressed as
\bea\lb{3.9}
{\bf H}&=&\frac1{\cal D}{(1+\beta{\bf D}^2+\kappa^2(2{\bf B}^2 +2\beta[{\bf B}\times{\bf D}]^2+\beta[{\bf B}\cdot{\bf D}]^2)+\kappa^4([{\bf B}\cdot{\bf D}]^2+[1+\beta{\bf D}^2]	{\bf B}^4 )} {\bf B}\nn\\
&&-\frac{\kappa^2}{\cal D} (1+\beta {\bf B}^2)(1+\kappa^2{\bf B}^2)({\bf B}\cdot{\bf D}){\bf D},
\eea
also in terms of the fields $\bf B$ and $\bf D$, in which the scalar quantity $\cal D$ in the denominator of \eq{3.9} reads
\be
{\cal D}={\sqrt{1+\beta {\bf B}^2}(1+\kappa^2{\bf B}^2)^{\frac32}\sqrt{1+\beta{\bf D}^2+\kappa^2{\bf B}^2+\beta\kappa^2({\bf B}\times{\bf D})^2}}.
\ee
Although \eq{3.9} is much more complicated than \eq{3.6},  we have the sharp asymptotic property
\be\lb{b3.10}
{\bf H}-{\bf B}=\mbox{O}(r^{-6}),\quad r=|x|\gg1,
\ee
similar to \eq{b3.7}, in view of \eq{b2.10} again.

Using \eq{b3.7}, \eq{b3.10}, and the method of the previous section, we obtain the exact results
\be\lb{b3.11}
Q_{f,e}=Q_e,\quad Q_{f,m}=Q_m,
\ee
as in the situation where $\kappa=0$. In other words, \eq{b3.11} holds for any $\kappa\geq0$, regardless of the local detailed behavior of $\bf E$ and $\bf H$.

\section{Finiteness of total energy in general situation}\lb{sec5}
\setcounter{equation}{0}

To compute the energy of such dyonic matter,  we note that the full Hamiltonian energy density is given by (cf. \cite{Yang3}, e.g.)
\bea\lb{3.13}
{\cal H}&=&\frac{{\bf E}^2+\kappa^2({\bf E}\cdot{\bf B})^2}{\sqrt{1-2\beta s}}-\frac1{\beta}\left(1-\sqrt{1-2\beta s}\right)\nn\\
&=&\frac{{\bf E}^2+\kappa^2({\bf E}\cdot{\bf B})^2}{\sqrt{1-2\beta s}(1+\sqrt{1-2\beta s})}+\frac{{\bf B}^2}{1+\sqrt{1-2\beta s}},
\eea
where the quantity $s$ is defined by \eq{1.2}. Thus, inserting \eq{3.3} and \eq{3.4} into \eq{1.2}, we have
\be\lb{3.14}
s=\frac{{\bf D}^2-{\bf B}^2+\kappa^2([{\bf B}\times{\bf D}]^2-{\bf B}^4)}{2(1+\beta{\bf D}^2+\kappa^2{\bf B}^2+\beta\kappa^2[{\bf B}\times{\bf D}]^2)}.
\ee
As a consequence of \eq{3.3}, \eq{3.4}, and \eq{3.14}, we see that \eq{3.13} becomes
\bea
{\cal H}&=&\frac{{\bf B}^2 {\cal R}_1{\cal R}_2+(1+\beta{\bf B}^2)({\bf D}^2+\kappa^2[{\bf B}\times{\bf D}]^2)}{{\cal R}_1({\cal R}_1+{\cal R}_2)},\lb{3.15}\\
{\cal R}_1&=&\sqrt{(1+\beta{\bf B}^2)(1+\kappa^2{\bf B}^2)},\quad{\cal R}_2=\sqrt{1+\beta{\bf D}^2+\kappa^2{\bf B}^2+\beta\kappa^2({\bf B}\times{\bf D})^2}.\lb{3.16}
\eea
Although this result reads complicated, it resumes a simple and symmetric form when $\kappa=0$:
\be\lb{3.24}
{\cal H}=\frac{{\bf B}^2\sqrt{1+\beta{\bf D}^2}+{\bf D}^2\sqrt{1+\beta{\bf B}^2}}{\sqrt{1+\beta{\bf B}^2}+\sqrt{1+\beta{\bf D}^2}}.
\ee
As a consequence of the asymptotic property \eq{b2.10} and \eq{3.15}--\eq{3.16}, we see that in all situations the Hamiltonian enjoys the behavior
\be\lb{b4.6}
{\cal H}=\mbox{O}(r^{-4}),\quad r=|x|\gg1,
\ee
such that the total energy of the system is always finite:
\be\lb{b4.7}
E=\int_{\bfR^3}{\cal H}\,\dd x<\infty.
\ee
The fine structure of \eq{3.15}--\eq{3.16} also indicates that the energy \eq{b4.7} depends on the detailed properties of the fields, and hence, those of the distributions of electric and
magnetic charges. Note that  \eq{b4.7} follows from \eq{b2.10} which ensures finiteness of electric and magnetic charges. In particular, we conclude that the energy of a system
of continuously distributed dyonic charges is finite, in all parameter regimes.

\section{Generalized formalism}\lb{sec6}
\setcounter{equation}{0}

Although our construction of the exact solutions in the previous sections is for the classical Born--Infeld theory consisting of \eq{1.1} and \eq{x2.2}, here we show that our method
works effectively in treating the generalized Born--Infeld theory of nonlinear
 electrodynamics model governed by the Lagrangian action density of  the following extended form:
\be\lb{4.1}
{\cal L}=f(s),
\ee
where $s$ is defined by \eq{x2.2} and $f(s)$ may simply be taken to be  a differentiable function fulfilling the normalization condition: 
\be\lb{x6.2}
f(0)=0,\quad f'(0)=1,
\ee
in order that the theory recovers the Maxwell theory in the weak-field limit. It is clear that the governing equations \eq{1.7}--\eq{1.6} are still valid but the 
classical constitutive equations \eq{1.3} and \eq{1.4} are updated by the generalized ones:
\bea
{\bf D}&=&f'(s)\left({\bf E}+{\kappa^2}({\bf E}\cdot{\bf B})\,{\bf B}\right), \lb{4.2} \\
{\bf H}&=&f'(s)\left({\bf B}-{\kappa^2}({\bf E}\cdot{\bf B})\,{\bf E}\right).\lb{4.3}
\eea
Accordingly, the nonlinear dielectrics and permeability coefficient matrix in \eq{1.9} is also modified:
\be
 \Sigma({\bf E},{\bf B})\equiv\left(\begin{array}{cc}f'(s)(1+\kappa^4({\bf E}\cdot{\bf B})^2)& \kappa^2({\bf E}\cdot{\bf B})\\ \kappa^2({\bf E}\cdot{\bf B})&\frac1{f'(s)}\end{array}\right).
\ee

We are to solve the system of the equations 
\eq{4.2} and \eq{4.3}. The key in the method used in the previous sections boils down to expressing the quantity $s$ given in \eq{x2.2} or \eq{1.2} in terms of the fields $\bf D$ and $\bf B$ which are determined by
solving the equations \eq{1.7} and \eq{1.8}.
To this end,
first we use
\eq{4.2} to arrive at the relations
\bea
 ({\bf B}\cdot{\bf D})^2&=&(f'(s))^2(1+\kappa^2{\bf B}^2)^2 ({\bf E}\cdot{\bf B})^2,\lb{5.30}\\
{\bf D}^2&=&(f'(s))^2 ({\bf E}^2+\kappa^2[2+\kappa^2{\bf B}^2][{\bf E}\cdot{\bf B}]^2).\lb{5.31}
\eea
Then, from \eq{5.30}, \eq{5.31}, and the Lagrange identity involving $\bf D$ and $\bf B$, we have
\bea\lb{5.32}
\frac{{\bf D}^2+\kappa^2({\bf B}\times{\bf D})^2}{1+\kappa^2{\bf B}^2}&=&(f'(s))^2({\bf E}^2+\kappa^2[{\bf E}\cdot{\bf B}]^2)\nn\\
&=&\left(f'\left(s\right)\right)^2\left(2s+{\bf B}^2\right).
\eea
Solving this implicit equation, we obtain
\be\lb{7.43}
s=s\left({\bf D}^2,{\bf B}^2,({\bf B}\times {\bf D})^2\right);\quad s\to0\mbox{ as }{\bf D},{\bf B}\to {\bf 0}.
\ee
Now multiplying \eq{4.2} by $\bf B$, we have
\be\lb{5.33}
{\bf E}\cdot{\bf B}=\frac{{\bf B}\cdot{\bf D}}{f'(s)(1+\kappa^2{\bf B}^2)}.
\ee
Thus, inserting \eq{5.33} into \eq{4.2}, we arrive at
\be\lb{5.34}
{\bf E}=\frac1{f'(s)}\left({\bf D}-\frac{\kappa^2({\bf B}\cdot{\bf D})}{1+\kappa^2{\bf B}^2}\,{\bf B}\right).
\ee
Furthermore, using \eq{5.33} and \eq{5.34} in \eq{4.3}, we obtain
\be\lb{5.35}
{\bf H}=-\frac{\kappa^2({\bf B}\cdot{\bf D})}{f'(s)(1+\kappa^2{\bf B}^2)}\,{\bf D}+\left(f'(s)+\frac{\kappa^4({\bf B}\cdot{\bf D})^2}{f'(s)(1+\kappa^2{\bf B}^2)^2}\right){\bf B}.
\ee

In view of  \eq{7.43} and \eq{b2.10}, we see that \eq{5.34} and \eq{5.35} satisfy \eq{b3.7} and \eq{b3.10}. Hence we arrive at \eq{b3.11} again.

Furthermore, the
 Hamiltonian energy density associated with the system reads \cite{Yang3,Yang-AOP}:
\bea\lb{b5.11}
{\cal H}&=&f'(s)({\bf E}^2+\kappa^2[{\bf E}\cdot{\bf B}]^2)-f(s)\nn\\
&=&\frac{{\bf D}^2+\kappa^2({\bf B}\times{\bf D})^2}{f'(s)(1+\kappa^2{\bf B}^2)}-f(s).
\eea
Using \eq{x6.2},  \eq{7.43}, and \eq{b2.10}  in \eq{5.32}, we have
\be\lb{b5.12}
s=\mbox{O}(r^{-4}),\quad f(s)=\mbox{O}(r^{-4}),\quad r=|x|\gg1.
\ee
Applying \eq{b5.12} in \eq{b5.11}, we arrive at \eq{b4.6} and \eq{b4.7} as before.

Summarizing, we have formally described a systematic method to construct the static solution of a finite energy to the full dyonic Born--Infeld nonlinear electrodynamics theory in its most general setting, in which
the key step is to solve for $s$ in the implicit equation \eq{5.32}, in terms of ${\bf D}^2, {\bf B}^2$, and $({\bf B}\times{\bf D})^2$, as stated in \eq{7.43}, such that the exact
solution to the full generalized Born--Infeld equations subject to the arbitrarily distributed continuous electric and magnetic charge densities $\rho_e$ and $\rho_m$
is expressed in terms of the fields ${\bf D},{\bf B},{\bf E},{\bf H}$ with $\bf E$ and $\bf H$ explicitly given by the formulas \eq{5.34} and \eq{5.35}, respectively.

\section{Some concrete model examples}\lb{sec7}
\setcounter{equation}{0}

In this section, we solve some concrete model examples to illustrate the effectiveness of our method. Following the formalism of Section \ref{sec6}, we know that the key in
the construction is to obtain the quantity $s$ as given in \eq{7.43}. Note that multicentered point-charge solutions are obtained in \cite{Yang-AOP}. Here we treat {\em continuous  situations}
and we emphasize that, in what follows, the equations are regarded as solved, in essence,  whenever $s$ is obtained in the form \eq{7.43}.

As a first  example of \eq{4.1}, consider the logarithmic model \cite{Soleng,Fe,AM,Gaete,K6}:
\be\lb{4.5}
f(s)=-\frac1\beta\ln(1-\beta s),\quad \beta>0.
\ee
Then \eq{5.32} assumes the form
\be\lb{x7.2}
\frac{{\bf D}^2+\kappa^2({\bf B}\times{\bf D})^2}{1+\kappa^2{\bf B}^2}=\frac{2s+{\bf B}^2}{(1-\beta s)^2}.
\ee
This is a quadratic equation in $s$ which can be solved to give us the result
\bea
s&=&\frac{\eta-{\bf B}^2}{1+\beta \eta+\sqrt{1+\beta\eta (2+\beta {\bf B}^2)}},\lb{4.27}\\
\eta&\equiv&\frac{{\bf D}^2+\kappa^2({\bf B}\times{\bf D})^2}{1+\kappa^2{\bf B}^2}.\lb{4.28}
\eea
For computational purposes, it is useful to note that
\be\lb{x7.5}
\eta={\bf D}^2 -\frac{\kappa^2 ({\bf B}\cdot{\bf D})^2}{1+\kappa^2{\bf B}^2}.
\ee

As a second example of \eq{4.1}, we consider the exponential model  \cite{H1,H2} defined by
\be\lb{5.1}
f(s)=\frac1\beta (\e^{\beta s}-1),\quad \beta>0.
\ee
Then \eq{5.32} becomes
\be
\frac{{\bf D}^2+\kappa^2({\bf B}\times{\bf D})^2}{1+\kappa^2{\bf B}^2}=\e^{2\beta s}(2s+{\bf B}^2).
\ee
For convenience, we now recast this equation into
\be\lb{6.7}
\beta\e^{\beta{\bf B}^2}\frac{({\bf D}^2+\kappa^2({\bf B}\times{\bf D})^2)}{1+\kappa^2{\bf B}^2}=\e^{2\beta s+\beta{\bf B}^2}(2\beta s+\beta{\bf B}^2),
\ee
which is of the Lambert $W$ function equation type, where the $W$ function \cite{CG,Chow}, $W(\tau)$, is defined by the equation
\be\lb{6.8}
\tau =W\e^W,
\ee
so that $W(\tau)$ is analytic in the interval $\tau>-\frac1\e$ and enjoys the asymptotic expansions
\be\lb{6.9}
W(\tau)=\left\{\begin{array}{ll}\sum_{k=1}^\infty \frac{(-k)^{k-1}}{k!} \tau^k,&  \tau \mbox{ is around }0,\\
&\\
                                               \ln \tau-\ln\ln \tau+\frac{\ln\ln \tau}{\ln \tau}+\cdots,& \tau>3,\end{array}\right.
\ee
where we have set in \eq{6.7} the quantities
\bea
W&=&2\beta s+\beta{\bf B}^2,\\
\tau&=&\beta\e^{\beta{\bf B}^2}\frac{({\bf D}^2+\kappa^2({\bf B}\times{\bf D})^2)}{1+\kappa^2{\bf B}^2}.
\eea
These results allow us to rewrite $s$ in terms of $\bf B$ and $\bf D$ as before:
\be\lb{5.13}
s= -\frac{{\bf B}^2}2+\frac1{2\beta}W\left(\beta \e^{\beta {\bf B}^2}\frac{({\bf D}^2+
\kappa^2[{\bf B}\times{\bf D}]^2)}{1+\kappa^2{\bf B}^2}\right),
\ee
which accomplishes our construction of the exact solution explicitly as described.

As a third example of \eq{4.1}, we  consider the quadratic model \cite{De,K2007,GS,C2015,K2017} defined by
\be\lb{5.18}
f(s)=s+\alpha s^2,
\ee
where $\alpha>0$ is a parameter. This is the simplest polynomial model  that excludes monopoles as finite-energy point-like magnetically charged particles \cite{Yang1,Yang4}
which are known to exist in the classical Born--Infeld theory.  Then \eq{5.32} reads
\be\lb{6.14}
\left(1+2\alpha s\right)^2\left(2s+{\bf B}^2\right)=\eta,
\ee
where $\eta$ is as defined by \eq{4.28} or given in \eq{x7.5}, which is a cubic equation in $s$. Solving this equation, we obtain
\be
s=-\frac1{3\alpha}-\frac{{\bf B}^2}6+\frac{\om^{\frac13}}{12\alpha}+\frac{1-\alpha{\bf B}^2(2-\alpha{\bf B}^2)}{3\alpha \,\om^{\frac13}},
\ee
where
\bea
\om&=&8\alpha^2{\bf B}^4(3-\alpha{\bf B}^2)+4(2+27\alpha \eta-6\alpha{\bf B}^2)+12{\gamma}^{\frac12},\\
\gamma&=&12\alpha\eta+\alpha^2\eta\left(81\eta-36{\bf B}^2+36\alpha{\bf B}^4-12\alpha^2{\bf B}^6\right).
\eea

Although these solutions appear complicated algebraically, their behavior in terms of the fields $\bf D$ and $\bf B$ is all clearly exhibited explicitly and exactly.

\section{Summary}\lb{sec8}

In this work, we have developed a systematic method to construct exact solutions to the governing equations of the classical Born--Infeld  electrodynamics theory and its generalizations  realizing arbitrarily distributed
continuous electric, magnetic, and dyonic charge sources, in static situations. The solutions are shown to carry finite energies as far as the prescribed electric, magnetic, and dyonic charges 
are finite which are consistently achieved by the Coulomb type asymptotic
behavior of the underlying electric and magnetic fields. It is also established that, in all situations, the induced free electric and magnetic charges coincide with the prescribed ones, which
extends the result of Born and Infeld in the singly point electric charge situation to the most general settings of arbitrary distributions of sources and arbitrary model nonlinearities.

\medskip

\medskip

{\bf Data availability statement}: The data that supports the findings of this study are
available within the article.

{\bf Conflict of interest statement}:
The author declares that he has no known competing financial interests or personal relationships that could have appeared to influence the work reported in this article.

\end{document}